\documentstyle[pra,preprint,eqsecnum,aps]{revtex}
\begin{document}
\draft
\title{
The Emergence of Classicality via Decoherence
Described by Lindblad Operators}
\author{Michael R. Gallis{\footnote{e-mail: mrg3@psuvm.psu.edu}}}
\address{
Department of Physics, Pennsylvania State University,
Schuylkill
Campus,\\
Schuylkill Haven, Pennsylvania 17972\\
}
\date{\today}
\maketitle
\begin{abstract}
Zurek, Habib and Paz
[W. H. Zurek, S. Habib and J. P. Paz, Phys. Rev. Lett. {\bf 70}
(1993)\ 1187]  have characterized the set of states of maximal
stability defined as the set of states having  minimum entropy
increase due to interaction with an environment, and shown that
coherent states are maximal for the particular environment model
examined.  To generalize these results, I consider entropy production
within the Lindblad theory of  open systems, treating environment
effects perturbatively. I characterize the maximally predicitive
states which emerge from several forms of effective dynamics,
including decoherence from spatially correlated noise.  Under a
variety of conditions,  coherent states emerge as the maximal states.
\end{abstract}
\pacs{PACS numbers 3.65.Bz, 3.65.-w, 5.40.+j}
\narrowtext

\section{Introduction}
\label{sec:intro}

Decoherence which results from a quantum system's interaction with
an environment can provide a mechanism for characterizing the
 transition from quantum to classical behavior for a quantum open
 system \cite{Joos&Zeh,Zurekold}, and has been an integral part of
 several programs addressing the emergence
of classicality \cite{Zurek,Halliwell}.  Zurek, Habib and Paz (ZHP)
 have characterized the effectiveness of decoherence in terms of a
 predictability sieve, and identified the maximally predictive states
(defined as those with minimal entropy production)
 as the most classical \cite{ZHP}.  ZHP considered an
environment model consisting of an independent
 oscillator bath linearly coupled to the system of
interest, which has been studied in the context of quantum
brownian motion \cite{CLplus}.
ZHP demonstrated for the high temperature limit of the environment
that the coherent states of a harmonic oscillator are
 maximally predictive, and that zero squeezing (corresponding to
coherent states) is maximal for squeezed states considered at
arbitrary environment temperatures.

The purpose of this paper is to extend the results of ZHP to
additional environment models.
The effects of the environment are considered in the general
framework of the Lindblad form for nonunitary evolution of a
harmonic oscillator \cite{Lindblad}, correponding to the Markov
limit.  In Sec.\ \ref{sec:perturbation}  I establish a systematic
framework for evaluating the predictability of states using first
order perturbation theory.  In  Sec.\ \ref{sec:linear}
I apply this framework to a family of Lindblad
generators which are at most quadratic in position and momentum whose
general properties have been studied in the literature
 \cite{Sandulescuplus}.  This family of Lindblad generators includes
as special cases  the Quantum Optical Master Equation \cite{gardiner}
and Dekker's model for Quantum dissipation \cite{Dekker}.  I consider
models with environment spatial correlation
effects \cite{localnoiseref,localdis} in Sec.\ \ref{sec:local}. I
comment on these results in Sec.\ \ref{sec:comments}.

\section{Perturbative Approach to Evaluating Predictability}
\label{sec:perturbation}

The predictability sieve was introduced by Zurek \cite{Zurek}
 as a means of characterizing those states which are most stable when
 considering not only the dynamics of the system, but also
 including the effects (such as decoherence) of interaction
 with the systems environment.  The set of  states having
 minimum (linear) entropy production are the best candidates
 for corresponding to points of classical phase space.  The
 linear entropy $\varsigma$ for a particular density operator
 $\rho$ is given by:
\begin{equation}
 \varsigma (t)={\rm Tr} [ \rho (t)
 -{ \rho (t)}^{ 2}]=1- {\rm Tr} [{ \rho (t)}^{ 2}].
\end{equation}
The effect of the environment on evolution is taken to be in the
Markov regime, and I assume that there is no further explicit time
 dependence on the Liouville generator for the isolated system.  The
 form of the evolution is then stationary, and we can write
\begin{equation}
\rho (t)={e}^{Lt}[\rho ] = J(t)[\rho].
\end{equation}
The generator $L$ is assumed to be comprised of two parts,
the evolution of the isolated system corresponding to $L_o$ with
$J_o(t) = e^{L_{o} t},$ and the
effects of the environment $\Delta L$.
I consider the effects of the environment perturbatively in part
because a wide variety of forms of evolution become tractable, and
also because many of the effective evolution equations are derived
using lowest order approximations such as the weak coupling
limit \cite{Alicki}.   From formal perturbation theory, the evolution
of Liouville generator can be written
\begin{eqnarray}
J(t) &=& J_o(t) + \int_{0}^{t}J_o(t-\tau)\Delta L J(\tau) d\tau
 \nonumber \\
&\cong& J_o(t) + \int_{0}^{t}J_o(t-\tau)\Delta L J_o(\tau) d\tau
+ O({\Delta L}^2)\nonumber
\\
&=& J_o(t) + J_1(t) +  O({\Delta L}^2),
\end{eqnarray}
defining $J_1(t)$ as the first order perturbation.  For an initial
 state
${\rho_o}$, the entropy at time t becomes \begin{eqnarray}
\varsigma (t) &=& 1-{\rm Tr}[J(t)[\rho_o] J(t)[\rho_o]]
 \nonumber \\
&\cong& 1-{\rm Tr}[J_o(t)[\rho_o] J_o(t)[\rho_o]]
 \nonumber \\
&&-2{\rm Tr}[J_o(t)[\rho_o] J_1(t)[\rho_o]].
\end{eqnarray}
If the isolated system has unitary evolution generated by a
hamiltonian $H$, then
\begin{eqnarray}
L_{o}[\rho] &=& {1 \over{i \hbar}} [H,\rho], \nonumber \\
J_o[\cdot] &=& U(t) \cdot U^{\dag}(t), \nonumber \\
U(t) &=& e^{{1 \over{i \hbar}} H t}.
\end{eqnarray}
In this case
\begin{equation}
1-{\rm Tr}[J_o(t)[\rho_o] J_o(t)[\rho_o]] =1-{\rm Tr}[\rho_o \rho_o]
= \varsigma(0).
\end{equation}
The maximal states are those with minimal entropy production,
that is, with minimized
\widetext
\begin{eqnarray}
\Delta \varsigma(t)&\equiv& -2{\rm Tr}[J_o(t)[\rho_o] J_1(t)[\rho_o]]
\nonumber \\
&=&-2\int_{0}^{t}\{{\rm Tr}[U(t)\rho_o U^{\dag}(t)
U(t-\tau)\Delta L[U(s)\rho_o U^{\dag}(s)]U^{\dag}(t-\tau)]\} d\tau
\nonumber \\
&=&-2\int_{0}^{t}{\rm Tr}[\rho_o
U^{\dag}(\tau)\Delta L[U(\tau)\rho_o U^{\dag}(\tau)]U(\tau)] d\tau,
\end{eqnarray}
\narrowtext
where the cyclic property of the trace and the unitarity of $U(t)$
have been used for the final simplification.

The general form for a Lindblad generator is
\begin{equation}
L[\rho ]={1 \over i\hbar}[H,\rho ]+{1 \over
2\hbar}\sum\nolimits\limits_{j } [{V}_{j }\rho
,{V}_{j}^{\dagger }]+[{V}_{j },\rho {V}_{j }^{\dagger }],
\end{equation}
so that the contribution to environment interaction can be
written as the perturbation
\begin{equation}
  \Delta L[\rho ]={1 \over i\hbar}[\Delta H,\rho ]+{1 \over
2\hbar}\sum\nolimits\limits_{j } [{V}_{j }\rho
,{V}_{j}^{\dagger }]+[{V}_{j },\rho {V}_{j }^{\dagger }].
\label{lindbladform}
\end{equation}
The entropy production can now be written
\begin{eqnarray}
\Delta \varsigma(t) &=&
-2\int_{0}^{t}{\rm Tr}[\rho_o
U^{\dag}(\tau){1 \over i\hbar}[\Delta H,U(\tau)\rho_o
U^{\dag}(\tau)]U(\tau) d\tau
\nonumber \\
&&-{1 \over \hbar}\int_{0}^{t}{\rm Tr}[\rho_o
U^{\dag}(\tau)
\sum\nolimits\limits_{j }
([{V}_{j }U(\tau)\rho_o U^{\dag}(\tau),{V}_{j}^{\dagger }]
\nonumber \\
&&+[{V}_{j },U(\tau)\rho_o U^{\dag}(\tau) {V}_{j }^{\dagger }]
)U(\tau)]d\tau.
\end{eqnarray}
The first term on the RHS of this equation is
identically zero, from the cyclic property of the trace.
The remainder can written in a simpler form using the
cyclic property of the trace, the unitarity of $U(t)$ and the
identification:
\begin{equation}
{V}_{j }(\tau) \equiv U^{\dag}(\tau){V}_{j }U(\tau)
\end{equation}
to yield
\begin{eqnarray}
\Delta \varsigma(t)&=&
{1 \over \hbar} \int_{0}^{t}\sum\nolimits\limits_{j }{\rm Tr}
[\{{V}_{j }^{\dag}(\tau){V}_{j }(\tau),\rho_o\}
\nonumber \\
&&-2{V}_{j }(\tau)\rho_o{V}_{j }^{\dag}(\tau)] d\tau.
\end{eqnarray}
If $\rho_o$ is a pure state, then it is also a projection, with
\begin{equation}
\rho_o = P = |\psi\rangle \langle \psi|
\end{equation}
with
\begin{equation}
P^2 = P
\end{equation}
and
\begin{equation}
P O P = P\langle \psi|O|\psi\rangle  =P \langle O\rangle
\end{equation}
for an arbitrary operator $O$.  Thus, for pure states, the entropy
production is given by
\begin{equation}
\Delta \varsigma(t)=
{2 \over \hbar} \int_{0}^{t}
\sum\nolimits\limits_{j }(
\langle \{{V}_{j }^{\dag}(\tau){V}_{j }(\tau)\rangle -
\langle {V}_{j }(\tau)\rangle \langle {V}_{j }^{\dag}(\tau)\rangle )
d\tau.
\label{entropy}
\end{equation}
Minimization of this final quantity can then be used to
determine the maximal states.

\section{Lindblad Operators Quadratic in Position and Momentum}
\label{sec:linear}

I will now apply the results of  Sec.\ \ref{sec:perturbation}
to Lindblad operators which have $\{V_{j}\}$ linear in
position and momentum.  This family of generators have been studied
extensively in the literature \cite{Sandulescuplus}, and include as
 special cases the quantum optical master equation \cite{gardiner}
 and Dekker's phenomenological master equation \cite{Dekker}.
In terms of Eq.\ (\ref{lindbladform}), the operators are given by
 \begin{eqnarray}
\Delta H&=& {\mu \over 2}\{x,p\}
\nonumber \\
{V}_{j }&=&{a}_{j }p+{b}_{j }x.
\end{eqnarray}
With the identifications
 \begin{eqnarray}
D_{qq}&=& {\hbar \over 2} \sum\nolimits\limits_{j } |a_{j}|^{2},
\nonumber \\
D_{pp}&=& {\hbar \over 2} \sum\nolimits\limits_{j } |b_{j}|^{2},
\nonumber \\
D_{pq}&=& {\hbar \over 2} \sum\nolimits\limits_{j }
-{\rm Re}[a_{j}b_{j}^{*}],
\nonumber \\
\lambda&=&  \sum\nolimits\limits_{j } {\rm Im}[a_{j}b_{j}^{*}],
\end{eqnarray}
the perturbation on the system evolution becomes
 \begin{eqnarray}
\Delta L [\rho]&=&{1 \over {i \hbar}}[{\mu \over 2}\{x,p\},\rho]
-{D_{qq} \over {\hbar^2}} [p,[p,\rho]]
-{D_{pp} \over {\hbar^2}} [x,[x,\rho]]
\nonumber \\
&&+{D_{pq} \over {\hbar^2}} ([x,[p,\rho]]+[p,[x,\rho]])
\nonumber \\
&&+{{i \lambda} \over {2 \hbar}} ([x,\{p,\rho\}]-[p,\{x,\rho\}]).
\label{evolution}
\end{eqnarray}
The particular choice of parameters $D_{qq},D_{pp},
D_{pq}, \lambda$ and $\mu$ Eq.\ (\ref{evolution}) determines
the details of the evolution (i.e. evolution corresponding to
 Quantum Optical Master Equation, Dekker's master equation,
etc.).  It has
been determined that only when \begin{equation}
{D_{pq}} = {\hbar \over 2} \sum\nolimits\limits_{j }
-{\rm Re}[a_{j}b_{j}^{*}] = 0
\end{equation}
will the system relax into a thermal equilibrium
state \cite{Sandulescuplus}, so
I adopt this condition for the remainder of the paper.

For the simple harmonic oscillator, the operator equations
of motion are easily solved by
\begin{eqnarray}
x(\tau)&=&U^{\dag}(\tau) x U(\tau)
\nonumber \\
&=&x \cos (\omega \tau) + {p \over {m \omega}}\sin (\omega \tau),
\nonumber \\
p(\tau)&=&U^{\dag}(\tau) p U(\tau)
\nonumber \\
&=&p \cos (\omega \tau) - { {m \omega} x}\sin (\omega \tau),
\label{SHO}
\end{eqnarray}
to yield
\begin{eqnarray}
{V}_{j}(\tau)&=&{a}_{j }(p \cos(\omega \tau)
-m\omega x \sin (\omega \tau))
\nonumber \\
&&+{b}_{j }(x\cos(\omega \tau)+{p \over {m \omega}}\sin
(\omega \tau)).
\label{vt}
\end{eqnarray}
Substituting Eq.\ (\ref{vt}) into Eq.\ (\ref{entropy})
and evaluating the elementary trigonometric integrals over $\tau$:
\begin{eqnarray}
\Delta \varsigma(t)&=&
f_{1}(t)
({1 \over {2m}}\langle p^{2}-\langle p\rangle ^{2}\rangle  +
{{m\omega^{2}}\over {2}}\langle x^{2} -\langle x\rangle ^{2}\rangle
)-2\lambda t \nonumber \\
&&+f_{2}(t)
({1 \over {2m}}\langle p^{2}-\langle p\rangle ^{2}\rangle  -
{{m\omega^{2}}\over {2}}\langle x^{2} -\langle x\rangle ^{2}\rangle )
\nonumber \\
&&+f_{3}(t)
({\omega \over 2}\langle \{x,p\}\rangle -\omega\langle x\rangle
\langle p\rangle),
 \label{entropy2}
\end{eqnarray}
where
\begin{eqnarray}
f_{1}(t)&=&
t2m({{2 D_{qq}}\over{\hbar^2}}+{{2 D_{pp}}\over{(m \omega
\hbar)^2}}),
\nonumber \\
f_{2}(t)&=&2m{{\sin (2\omega t)}\over{2 \omega}}
({{2 D_{qq}}\over{\hbar^2}}-{{2 D_{pp}}\over{(m \omega \hbar)^2}}),
\nonumber \\
f_{3}(t)&=&-2{{\sin^{2}(\omega t)}\over{\omega^2}}
({{2 D_{qq}}\over{\hbar^2}}-{{2 D_{pp}}\over{(m \omega \hbar)^2}}).
\end{eqnarray}

Eq.\ (\ref{entropy2}) is the expectation of a c-number times the
 harmonic oscillator hamiltonian plus a second c-number constant
after sqeezeing and translation by $\langle x\rangle $ in position
 and $\langle p\rangle $ in
momentum.  The state which minimizes Eq.\ (\ref{entropy2}) will be
the corresponding squeezed  and translated ground state, which is
simply a coherent squeezed state \cite{Hirota}.  In terms
 of the harmonic oscillator
creation and annihilation operators $a^{\dag}$ and $a$,
the squeeze operator is given by
\begin{equation}
S(\zeta)=e^{{1\over 2}(\zeta^{*}a^{2}-\zeta a^{\dag 2})},
\end{equation}
 and Glauber's displacement operator is given by
\begin{equation}
D(\alpha)=e^{(\alpha a^{\dag}-\alpha a)}.
\end{equation}
The appropriate selection of the paramter $\alpha$ for the
displacement operator
\begin{eqnarray}
{\rm Im}(\alpha)&=& ({\hbar\over{2m\omega}})^{1\over 2} \langle
p\rangle,  \nonumber \\
{\rm Re}(\alpha)&=& ({{\hbar m\omega}\over 2})^{1\over 2} \langle
x\rangle,  \end{eqnarray}
provides the necessary translation (in phase space).  For simplicity,
I take   $\langle x\rangle=0$ and $\langle p\rangle = 0$ for
the rest of this section.
The effect of squeezing on the anihillation operator can be written
\begin{equation}
S^{\dag}(\zeta)aS(\zeta)=\mu a + \nu a^{\dag},
\end{equation}
where
\begin{eqnarray}
\zeta&=&s e^{i \theta},
\nonumber \\
\mu&=&\cosh (s),\nonumber \\
\nu&=&\sinh (s) e^{i \theta}.
\end{eqnarray}
The parameter $s$ determines the amount of squeezing
($s=0$ corresponding to no squeezing), and the parameter
$\theta$ determines the orientation of the sqeeze axis.
The squeezed harmonic oscillator hamiltonian is given by
\begin{eqnarray}
&&S^{\dag}(\zeta)({p^{2} \over {2m}}+{m\omega^{2}
\over 2}x^{2})S(\zeta)
\nonumber \\
&=&\cosh (2s)
({m \omega^{2} \over 2}x^{2}+{p^{2} \over {2m}})
\nonumber \\
&&+\sinh (2s)\sin (\theta)
({m \omega^{2} \over 2}x^{2}-{p^{2} \over {2m}})
\nonumber \\
&&+\cosh (2s)\cos (\theta){\omega \over 2}
\{x,p\},
\end{eqnarray}
which can be used to identify the amount and direction of squeezing
required to minimize
Eq.\ (\ref{entropy2}).  The easiest condition to extract is the
direction of squeezing:
\begin{equation}
\cos ( \theta ) = {f_{3}(t)\over f_{1}(t)}.
\end{equation}
Since $f_{1}(t)$ is linear in $t$ while $f_{3}(t)$ oscillates, the
long time behavior is $\theta = 0$. The amount of
squeezing can be determined by examining
\begin{equation}
(1+ \cosh^{2}(2s))\sin(\theta)={f_{2}(t)\over f_{1}(t)}\cosh(2s).
\label{sqlimit}
\end{equation}
To analyze the long time behavior, it is useful to note that
$\sin(\theta)$ approaches $1$ and ${f_{2}(t)/f_{1}(t)}$ approaches
$0$ (and therefore so does the RHS of Eq.\ (\ref{sqlimit})).
Thus, the long time behavior of the amount of squeezing required
for minimum entropy production is
\begin{equation}
(1+ \cosh^{2}(2s))=0,
\end{equation}
which implies $s=0$, no squeezing.
 Thus, for times on the order of the dynamical time scale of the
 system (more than a few cycles), coherent states are the maximal
states, just as ZHP found for their environment model.

\section{Correlation Effects in environment noise}
\label{sec:local}
In this section I wish to consider environment models which include
the effects of finite correlation lengths in the environment
\cite{localnoiseref,localdis}.
{}From a strictly phenomenological point of view, Quantum Mechanics
 with Spontaneous Localization  can be included by virtue of the
effective form of the dynamics, although this
is actually a fundamental modification of Quantum
Mechanics \cite{GRW}. Many of these models can be written in the form
\begin{eqnarray}
{ \partial \rho (x,x';t) \over \partial t}&=&{\rm Hamiltonian}
+{\rm Dissipation}
\nonumber \\
&&+\cdots-g(x,x')\rho (x,x';t),
\label{evolutionlocal}
\end{eqnarray}
where the decoherence term which we shall focus upon
can be expressed in terms of the correlations of
a classical fluctuating potential $V({ x},t)$ with
\begin{equation}
g(x;y) = {1\over{\hbar^2}}(c(x;x)+c(y;y)-2c(x;y)),
\end{equation}
where
\begin{equation}
\langle  V({ x},t)V({ y},\tau)\rangle_{\rm av} =c({ x};{y})\delta
(t-\tau). \label{cor}
\end{equation}
For simplicity, I consider only a homogeneous and isotropic
environment for which $c({ x};{y})=c({ x}-{y})$, and
$g({ x};{y})=g({ x}-{y})$.
I will also restrict my attention to weak dissipation , and
consider only the evolution due to the unperturbed hamiltonian
 and the
(spatially correlated) noise term.

The Lindblad form can be used to represent the noise
term with
\begin{equation}
\{V_j\}=a(k)e^{i kx}
\label{planewave}
\end{equation}
and replacing the discrete sum over $j$ in
Eq.\ (\ref{lindbladform}) with an integral over
$k$.  The evolution in this case can be written
\widetext
\begin{equation}
{ \partial \rho (x,x';t) \over \partial t}={\rm Hamiltonian}
-{1 \over \hbar}\int^{}_{}dk |a(k)|^{2}(1-e^{i k (x
-x')})\rho (x,x';t),
\end{equation}
\narrowtext
so that $|a(k)|^{2}$ and $c(r)$ are Fourier transform pairs:
\begin{equation}
c(r) = {{\hbar} \over 2}
\int_{}^{}|a(k)|^{2}e^{ik(r)} dk.
\label{correlation}
\end{equation}
Thus
a noise term with short correlation length scales will
have a narrow $c(r)$ and a broad $|a(k)|^{2}$ while
a long correlation length scale implies a narrow $|a(k)|^{2}$.
Inserting  Eq.\ (\ref{planewave}) into Eq.\ (\ref{entropy})
yields
\begin{equation}
\Delta \varsigma(t)=
{1 \over \hbar} \int_{0}^{t}
\int_{}^{}|a(k)|^{2}(1-|\langle e^{ikx(\tau)}\rangle |^{2})dk d\tau,
\label{entropylocal}
\end{equation}
where $x(\tau)$ is given by  Eq.\ (\ref{SHO}),
and $e^{ikx(s)}$ can immediately be recognized as Glauber's
displacement operator, with a translation in momentum
of $\hbar k \cos (\omega \tau)$ and a translation in position
of $-{(\hbar k)/(m \omega)} \sin (\omega \tau)$.
In terms of the translated state
\begin{equation}
|\psi;k,\tau\rangle \equiv e^{ikx(\tau)}|\psi\rangle ,
\end{equation}
the entropy production is given by
\begin{equation}
\Delta \varsigma(t)=
{1 \over \hbar} \int_{0}^{t}
\int_{}^{}|a(k)|^{2}(1-|\langle \psi|\psi;k,\tau\rangle |^{2})dk
d\tau. \label{entropylocal2}
\end{equation}

While it is not possible to find general solutions for
 the minimization of
Eq.\ (\ref{entropylocal2}) for arbitrary environment correlations,
it is possible to extract important limiting cases.  Restricting
attention to times of several oscillator periods or more,
entropy minimization requires the {\em maximization} of
$|\langle \psi|\psi;k,\tau\rangle |^{2}$ for typical values of $k$,
on the order of
$\delta k$ (the spread of $|a(k)|^{2}$), resulting in typical
translations of $\hbar \delta k$ in momentum and
${(\hbar \delta k)/(m \omega)}$ in position.
The maximization of the square of the inner product of any
 two normalized vectors occurs when the vectors are identical
(up to a phase).  Thus the maximal states and the translated maximal
states will be approximately equal for typical translations,
requiring that the width of  $|a(k)|^{2}$ be much less than
the width of the maximal states:
\begin{eqnarray}
\delta k &\ll& {{m\omega}\over\hbar}\Delta x
\nonumber \\
\delta k &\ll& {{1}\over\hbar}\Delta p.
\label{width1}
\end{eqnarray}
Since $|a(k)|^{2}$ and $c(r)$ are Fourier transform pairs,
this last condition also implies that the noise
spatial correlation function $c(r)$ is wide (compared to the
maximal states).  In this long correlation length scale limit,
$g(x,x')$ in  Eq.\ (\ref{evolutionlocal}) is quadratic
\cite{localnoiseref}, corresponding
to the low dissipation limit studied by ZHP and to the results
of Sec.\ \ref{sec:linear} with $b_{j} = 0$ for all $j$.  The coherent
states are then the maximal states if they are consistent with the
condition expressed in Eq.\ (\ref{width1}) using coherent state
values for $\Delta x$ and $\Delta p$:
\begin{equation}
\delta k \ll \sqrt{{m\omega}\over{2\hbar}},
\end{equation}
which corresponds to a environment
corellatation length much larger than the width of a coherent state
$\Delta x = \sqrt{\hbar/2m\omega}$.

If the environment correlation length is shorter than the width
of the coherent state, then the approximation described above
is not valid.  It is useful to examine the entropy production in
the position representation and absorbing the time dependance into
the (Schr\"{o}dinger picture) state vectors, where
\begin{eqnarray}
\langle \psi|e^{ikx(\tau)}|\psi\rangle &=& \langle
\psi|U^{\dag}(\tau)e^{ikx}U(\tau)|\psi\rangle  \nonumber \\
&=& \langle \psi(\tau)|e^{ikx}|\psi(\tau)\rangle  \nonumber \\
&=&\int_{}^{} dx e^{ikx} |\psi(x,\tau)|^{2} \nonumber \\
&=& \int_{}^{} dx e^{ikx} P(x,\tau).
\end{eqnarray}
Entropy production then becomes
\begin{equation}
\Delta \varsigma(t)=
{2 \over {\hbar ^2}}[c(0) t -\int_{}^{} d\tau dx dx' c(x-x')
 P(x,\tau) P(x',\tau)],
\label{entropypos}
\end{equation}
using the Fourier transform relation expressed in
Eq.\ (\ref{correlation}).  For narrow $c(r)$ this expression becomes
independent of the function $P$.  To see this, consider an example
 where the spatial correlation of the environment is given by
\begin{equation}
c(r) =
\lambda e^{({r \over \sigma})^{2}}.
\end{equation}
In the limit $\sigma$ approaches zero
\begin{eqnarray}
\int_{}^{} dx dx' \lambda e^{({{x-x'} \over \sigma})^{2}}
 P(x,\tau) P(x',\tau) &\cong&
\nonumber \\
\lambda \sigma \sqrt{\pi} \int_{}^{} P^{2}(x,\tau) dx&\rightarrow& 0,
\end{eqnarray}
so that in the short correlation length limit
\begin{equation}
\Delta \varsigma(t)=
{2 \over {\hbar ^2}}c(0)t
\label{increaselimit}
\end{equation}
so that in this regime, all states produce the same entropy,
there are no maximal states.  However, the decay rate of the
off diagonal terms given by the decoherence term in
Eq.\ (\ref{evolutionlocal}) in the low short correlation
length regime is generally at a maximum,
\begin{equation}
{\partial \rho (x,x';t) \over \partial t}=
\cdots-{{2c(0)}\over {\hbar ^2}}\rho (x,x';t),
\end{equation}
and if decoherence is to be effective, the decay
time must be comparable to dynamical timescales.
There will necessarily be a significant increase in the entropy
for all pure states, so that all states will be rapidly \`` mixed\''
by the noise.

\section{Comments and Conclusions}
\label{sec:comments}

I have established an approximation scheme for determining maximal
states (as defined by ZHP), and applied it to two families of
Lindblad operators. For Lindblad operators which are
at most quadratic in position and momentum, squeezed states emerge
as the maximal states for intermediate times compared to the
dynamical time scales.  The amount of squeezing decreases with time,
so that coherent states are maximal for large timescales.  Large
 timescales are the most relevant, since an object's classicality
should be an enduring property, not a transient one.  For an
 environment with finite spatial correlation, coherent states
emerge as maximal when the environment has long correlation length,
but all states  rapidly become mixed states when the environment
length scale is long.  Thus environment correlation effects will not
be important in establishing the nature of maximal states and the
character of quasiclassical states.  However, correlation effects
can still be important when considering quantum interference between
two such states.  One important result which emerges from these
 calculations is that coherent states are a robust choice
for the maximal states.

\end{document}